\documentclass[letterpaper]{article} 

\usepackage{aaai25} 
\usepackage{times}  
\usepackage{helvet}  
\usepackage{courier}  
\usepackage[hyphens]{url}  
\usepackage{graphicx} 
\usepackage{natbib}  
\usepackage{caption} 
\usepackage{algorithm}
\usepackage{algorithmic}

\usepackage{array}

\usepackage{newfloat}
\usepackage{listings}
\DeclareCaptionStyle{ruled}{labelfont=normalfont,labelsep=colon,strut=off} 
\floatstyle{ruled}
\newfloat{listing}{tb}{lst}{}
\floatname{listing}{Listing}
\title{Mapping the Regulatory Learning Space for the EU AI Act}

\author {
    Dave Lewis\textsuperscript{\rm 1},
    Marta Lasek-Markey\textsuperscript{\rm 1},
    Delaram Golpayegani\textsuperscript{\rm 1},
    Harshvardhan J Pandit\textsuperscript{\rm 2}    
}
\affiliations {
    \textsuperscript{\rm 1}ADAPT Centre at Trinity College Dublin\\
    \textsuperscript{\rm 2}AI Accountability Lab at Trinity College Dublin\\
    delewis@tcd.ie, mlasek@tcd.ie, golpayes@tcd.ie, me@harshp.com
}

\usepackage{bibentry}

\begin{document}

\maketitle

\begin{abstract}

The EU AI Act represents the world’s first transnational AI regulation with concrete enforcement measures. It builds on existing EU mechanisms for regulating health and safety of products but extends them to protect fundamental rights and to address AI as a horizontal technology across multiple application sectors. We argue that this will lead to multiple uncertainties in the enforcement of the AI Act, which coupled with the fast-changing nature of AI technology, will require a strong emphasis on comprehensive and rapid regulatory learning for the Act. We define a \textit{parametrised regulatory learning space} based on the provisions of the Act and describe a layered system of different learning arenas where the population of oversight authorities, value chain participants, and affected stakeholders may interact to apply and learn from technical, organisational and legal implementation measures. We conclude by exploring how existing open data policies and practices in the EU can be adapted to support rapid and effective regulatory learning.
\end{abstract}

%
\section{Introduction}
Advances in research on fairness, accountability and transparency of algorithms has normalised approaches towards achieving ethical and trustworthy AI in response to societal concerns and risks posed. Where recent surveys regarding trustworthy AI guidelines demonstrate a high degree of consensus on common principles, including human rights and agency, privacy, non-discrimination, safety and robustness, transparency and accountability~\cite{hagendorff_ethics_2020, jobin_global_2019,fjeld_principled_2020, correa_worldwide_2023}, others argue that such a consensus on principles is not reflected in their implementation and raises new forms of avoidance. For instance, using techno-centric approaches such as differential privacy and explainable AI to avoid external inspection of AI design decisions and association with ethical dilemmas and affected stakeholders~\cite{palladino_biased_2023}, or where such principles often represent ``ethics washing'' as a distraction from harmful behaviour \cite{schultz_digital_2024}. In addition, uncertainty about the implementation of ethical AI principles is magnified by the rate at which new risks emerge due to the accelerating pace of technological advancement and adoption such as with Large Language Models (LLMs) and Generative AI (GenAI) \cite{grace_thousands_2024}.  

The AI Act, first proposed in April 2021 and which entered into force in August 2024, makes the EU the first jurisdiction to regulate AI technologies~\cite{noauthor_regulation_2024}. The Act takes a risk-based approach modelled on the existing EU scheme, known as the New Legislative Framework (NLF), for harmonising product health and safety legislation across the EU internal market  \cite{mazzini2023proposal}. Thus, the AI Act benefits from established rules and practices for product regulation laid out in the so-called ``Blue Guide''~\cite{eu-nlf}. The legislative process preceding the adoption of the Act eloquently demonstrated the technology pacing problem. As powerful classes of General Purpose AI (GPAI) threatened to evade enforcement under the NLF, which is predicated on products with a specific purpose, new provisions had to be introduced to handle the parallel advances in LLMs and GenAI. Market actors, regulators and civil society now have a concrete timetable within which the many remaining uncertainties in enforcing provisions of the Act present a substantial challenge \cite{kaizenner_ai_2024,keber_eu_2024}.  
 

In this paper, we focus on two major areas of uncertainty arising from novelties that the AI Act brings to the NLF. Firstly, the AI Act expands the scope seen in current legislation harmonised under the NLF from protection of health and safety to include protection of fundamental rights. Secondly, the AI Act attempts to regulate AI as a horizontal technology using enforcement mechanisms that are conducted on a sectoral basis under the NLF,  i.e. for specific classes of products such as the Medical Device Regulation (MDR)~\cite{mdr}.

We contend that given the regulatory uncertainties, pace of technological progress, and the relative immaturity of AI risk assessments and mitigations, the Act should in the first instance serve as a framework for learning the appropriate technical, organisational and regulatory measures which then shape future regulatory and technical progress. We therefore argue that a key benefit of the AI Act is the motivation to mobilise resources to resolve and build consensus on these uncertainties in a way that addresses existing and future AI risks in an efficient and legitimate manner. We phrase this gain of knowledge as `learnings'.

To identify such learnings, we analyse the primary uncertainties arising from the AI Act and define `learning arenas' that are useful to structure and maximise the shared efficiency of learnings under the current complex structures defined in the Act. We conclude with concrete recommendations on how to improve the information flow between actors across learning arenas for accelerating shared learning and to better respond to the pace of advances in AI.

\section{Fundamental Rights and Legal Uncertainties}
This paper argues that regulatory learning in the context of the AI Act should begin, at the most general level, by adequately identifying the role of fundamental rights and addressing issues related to the so-called direct effect of EU law. This section will briefly outline the three sources of fundamental rights protection in the EU with particular focus on issues with direct effect of the EU Charter of Fundamental Rights (CFR). It will then consider the application of the Charter to the AI Act.

As EU integration progressed over the years, so did protection of fundamental rights. Consequently, within the EU legal order, there is currently no single uniform source of fundamental rights protection ~\cite{spaventa2018should}. Instead, Article 6 TEU identifies three sources thereof, in chronological order: 1) constitutional traditions shared by the Member States; 2) the European Convention for the Protection of Human Rights and Fundamental Freedoms (ECHR), which all EU Member States are parties to but the European Union is not; and 3) the EU Charter of Fundamental Rights. The latter has been gaining in importance since the Treaty of Lisbon's entry into force (2009), however, the scope of application of the Charter is limited by its so-called horizontal provisions, particularly Article 51. With Lenaerts, we note that the CFR rights are ``addressed to the institutions, bodies, offices and agencies of the Union with due regard for the principle of subsidiarity and to the Member States only when they are implementing Union law''~\cite{lenaerts2012exploring}. Therefore, while the provisions of the Charter bind the EU itself and the Member States, and this has been particularly evident in policymaking, its impact in litigation has been limited due to issues with direct effect of the Charter~\cite{frantziou2015horizontal, frantziou2019horizontal}. To be capable of direct effect, i.e. to create rights for individuals, a provision of EU law must fulfil the following three conditions laid down in the CJEU case 26/62 \textit{Van Gend en Loos}: it must be 1) clear, 2) precise and 3) unconditional and, therefore, not require any further implementing measures.

In relation to CFR rights, the CJEU has to date only considered direct effect of four provisions. Consequently, Article 21 CFR (right to non-discrimination), Article 31(2) CFR (right to maximum working hours, to daily and weekly rest periods and to annual leave) and Article 47 CFR (right to an effective remedy and to a fair trail) have been deemed as having direct effect~\cite{prechal2020horizontal}. Conversely, Article 27 CFR (workers' right to information and consultation) has been denied direct effect in case C‑176/12 \textit{AMS}, where the CJEU stated, in para 45, that ``for this article to be fully effective, it must be given more specific expression in European Union or national law''. Therefore, it does require additional implementing measures. Regarding the remainder of the CFR, it is not clear which of the rights might be capable of direct effect and, thus, whether individuals might be able to successfully rely on them.

Consequently, it is possible that in conflicts involving fundamental rights in the context of the AI Act, the other sources of protection, i.e. the European Convention on Human Rights which is binding on all the EU Member States, and indeed the national constitutions of the Member States, might also become relevant. As the EU and the Council of Europe are in the process of negotiating EU's accession to the ECHR, a rebuttable presumption of equivalent protection between the two systems, laid down in Article 52(3) of the CFR, continues to apply ~\cite{de2012doctrine}. Indeed, the European Court of Human Rights (ECtHR) has delivered judgments where violations of the Convention were found, e.g. in relation to unjustified processing of biometric data (case \textit{Glukhin v Russia}, Application no. 11519/20). In this vein, the body of ECtHR case law constitutes yet another important resource for regulatory learning pertaining to fundamental rights.  

While potentially all fundamental rights, including CFR rights, might fall within the scope of application of the AI Act, the Act explicitly refers to 17 fundamental rights in Recital 48 of the Preamble, which had also been mentioned in the Proposal for the AI Act (COM(2021) 206 final). This implies that the European Commission expects these to be of particular relevance in the context of AI, even though it does not grant them any additional legal standing. 
It is unclear which of these fundamental rights beyond the right to non-discrimination, the right to an effective remedy and to a fair trial, as well the specific workers' right to rest, might have direct effect. The lack of direct effect might constitute a barrier to effective enforcement of fundamental rights, particularly in horizontal disputes involving non-state actors, i.e. private parties~\cite{viljanen2023horizontal, fornasier2015impact}. In the context of the AI Act, disputes involving fundamental rights might potentially concern such types of deployers of high-risk AI systems as private employers, insurance companies or banks. Regarding fundamental rights that do not have direct effect, e.g. the workers' right to information and consultation, it appears that such non-state actors would not be bound by the Charter. 

Many of the fundamental rights enshrined in the Charter have further been concretised in secondary EU legislation, some of which also has direct effect and would, therefore, be binding on private parties. A prime example is the right to the protection of personal data guaranteed by the General Data Protection Regulation 2016/679 (GDPR), referenced multiple times across the AI Act, as well as other legislation, such as the Law Enforcement Directive 2016/680 or the ePrivacy Directive 2002/58. Other examples of secondary legislation mentioned in the AI Act which is connected to various CFR rights include Regulation 2024/900 on the transparency and targeting of political advertising; Framework Directive 2002/14 on employee information and consultation or the Platform Workers Directive 2024/2831. Additionally, the entire body of EU consumer legislation gives effect to Article 38 of the Charter (the right to consumer protection), including the Medical Devices Regulation 2017/745, the Representative Actions Directive 2020/1828 and the Unfair Commercial Practices Directive 2005/29 to name a few examples mentioned in the AI Act. Furthermore, there is other relevant EU secondary legislation which, albeit not explicitly referenced in the AI Act, provides additional layers of protection of various Charter rights. A pertinent example is the EU non-discrimination law consisting of the Framework Directive 2000/78, the Racial Equality Directive 2000/43 and the Gender Recast Directive 2006/54, where implications of AI and algorithmic management in employment are significant~\cite{wachter2021fairness, xenidis2020eu}.   
Given the existing regulatory complexities and uncertainties surrounding the status of fundamental rights and the Charter in the context of the AI Act, the following list of questions need to be prioritised towards ensuring predictable and consistent enforcement of the Act:
\begin{enumerate}
  \item Will all rights under the CFR be \textit{equally} protected?
  \item What role will Market Surveillance Authorities (MSAs) in relation to identifying whether a high-risk AI system has breached fundamental rights protections and enforcement of corrective measures?
  \item How will the enforcement powers of MSAs in enforcing rights protections (executed in concert with national bodies protecting fundamental rights as per Article 77 of the AI Act) interplay with those of notified bodies, national competent authorities?
  \item What role and weight will be placed on the Fundamental Rights Impact Assessment (FRIA) conducted by AI deployers and AI providers (Article 27) in enforcing fundamental rights protections?
\end{enumerate}

\section{Integrating Horizontal \& Vertical Requirements}
The AI Act takes a risk-based approach with technical requirements that fall primarily on \textit{AI providers} that aim to place an AI system on the market or into service in the EU, as well as organisations that deploy such AI offerings, termed \textit{AI deployers}. The requirements specify the ex-ante application of quality management (Article 17) and risk management techniques (Article 9) and involve technical measures for data governance and data protection (Article 10), technical documentation and record keeping (Articles 11 and 12), transparency (Article 13) and human oversight in AI system operation (Article 14), and implementation of accuracy, robustness and cybersecurity measures (Article 15). While the requirements are similar in scope to many existing voluntary ethical AI guidelines, their integration into the  NLF  poses important questions. In particular, it remains to be seen whether satisfying these requirements will adequately deliver the Act's goals of a well-functioning internal market balanced with high levels of protections for health, safety and fundamental rights (Recital 1) \cite{evers_talking_2024}. 

Adopting the well-established pattern of the NLF, the enforcement of technical requirements combines ex-ante certification of AI-based products classified as ``high-risk'' prior to release and ex-post monitoring and surveillance of such products once released and in use. The required certification process depends on the class of the high-risk AI system in question, of which the Act defines two sets. The first relates to the use of AI in products that are already subject to harmonised product legislation and listed in Annex I of the Act, which would receive certification on conformance to the Act via the certification mechanisms established therein. The second relates to the use of AI in applications not subject to existing product legislation but are deemed to represent risks needing oversight and listed in Annex III of the Act, for which self-certification of compliance will suffice, apart from the area of remote biometric identification (Annex III, point 1) for which a new certification mechanism will be established. The ex-post enforcement of requirements on high-risk AI systems involves planned post-market monitoring of deployed systems by providers, adherence by deployers to the instructions of use provided with the system and surveillance of these by a network of Market Surveillance Authorities (MSA). If an MSA has reason to believe that a high-risk AI system placed on the market does not comply with the Act's requirements, it has the power to obtain detailed technical documentation and resources from the provider, up to and including copies of the system code in order to make determination of breaches of requirements. Depending on the severity of any assessed breaches, an MSA may require corrective measures to be made to the system or its instructions for use or that the system is removed from the market. It may also impose financial penalties on the provider or deployer. 

The ability to unambiguously verify that a high-risk AI system employs appropriate technical measures to mitigate risks to health, safety and fundamental rights relies on investigation of technical documentation, testing and instructions, which therefore are central to the consistent and effective implementation of the Act, and where efficiency is vital to minimising the regulatory burden of complying with the Act~\cite{golpayegani2024AIcards}. However, the Act is limited to essential high-level requirements for technical measures and how they should be undertaken is stated in horizontal terms independent of sectoral context. Instead, in common with some other regulations under the NLF, the Act allows compliance with requirements to be demonstrated by provider in terms of a presumption of conformity gained by adherence to separate technical standards. These technical standards can be either ``Harmonised Standards'' developed by European Standards Organisations (ESOs) or if that is not possible or insufficient then via ``Common Specification'' to be developed by the European Commission (EC) \cite{tartaro2023towards}. 
The EC has already issued a request for harmonised standards in 10 areas to the ESOs, addressing different aspects of the required technical measures \cite{standardisationrequest}. A Joint Technical Committee of the ESO's CEN and CENELEC (JTC21) has been chartered in response to develop European Standards for AI. 

By developing the technical specifications needed to implement the Act's technical provisions in this way, ESOs can leverage expertise of industry practitioners, as well as draw upon existing international standards so that resulting specifications are technically correct, reflective of the recent state of the art and more readily implementable by industry actors. However, this presents several additional uncertainties:
\begin{itemize}
    \item \textit{EU-wide coordination of national regulatory authorities}: Under the NLF, product legislation is enforced at the national level, primarily by relevant MSAs, where legislations in Annex I of the Act involve such MSAs when products in their competent vertical sectors involve AI. However, MSAs in each of these regulated sectors are already part of an EU wide Administrative Coordination Group (AdCos) to coordinate and ensure consistent enforcement, and where the AdCos groups in turn coordinate on common NLF implementation issues.

    It is unclear if similar EU wide coordination is envisaged for each of the separate areas introduced to the NLF under Annex III of the AI Act, or if this will be coordinated as a group via a network of the AI National Competent Authorities (NCA). Further, it is unclear if issues raised under Annex I sectoral enforcement will be escalated to the AI Office and Board at EU level by that sector's Administrative Coordination Group or via the NCA. Similarly, for consistency of GDPR, EU-wide coordination happens through the European Data Protection Board (EDPB) whose members are supervisory authorities. The AI Act is unclear on whether the similarly defined European Artificial Intelligence Board (EAIB) in Article 65, will involve EU-wide coordination bodies related to fundamental rights directly, or whether this is to occur via the NCA. For example, in Ireland the national competent authority for the AI Act is yet to be announced while there is a notified body already operating to serve the certification needs of the large medical device sector, and in addition there are 10 different agencies and departments covering the MSA role for the 20 Annex I legislative acts as well as 9 fundamental rights protection bodies. Therefore, the EU-wide coordination to implement the AI Act involves numerous entities and is thus considerable. 

    \item \textit{Limited representation in standards development:} Standards development is a meticulous, lengthy and highly technical process. Therefore, the bulk of participation comes from large multinational companies that sustain teams of suitably-experienced experts, whereas other less well-resourced experts from NGOs, SMEs or academia struggle to maintain effective level of contribution \cite{cuccuru_interest_2019}. This raises concerns that the resulting AI harmonised standards from CEN/CENELEC JTC21 may not sufficiently represent the interests of citizens and SMEs in favour of the large companies that the Act must also regulate \cite{veale_demystifying_2021}. Further, JTC21 is able to adopt standards directly from ISO/IEC, where the subcommittee on AI (SC42) has been developing AI standards since 2018, many of which align to topics in the EC's harmonised standards request, e.g. ISO/IEC 23894 on AI risk management \cite{noauthor_isoiec_nodate-6} and ISO/IEC 24029-2 on AI robustness \cite{noauthor_isoiec_nodate-5}. However, as these are developed by experts from a global membership of national standards bodies rather than just the European-based ones, the representation of \textit{weaker} European actors may be further diluted. 
 
    \item \textit{Lack of free access to harmonised standards:} Standards bodies such as ISO/IEC and CEN/CENELEC rely on charging for access as means of funding to maintain their convening and quality assurance functions, and harmonised standards are typically high-demand items. However, as harmonised standards essentially enter EU law and are key to enforcing important consumer protections, paywall access became untenable and resulted in court actions \cite{kamara_general_2022}. The European Court of Justice delivered a judgment in Case C-588/21 P that there was an overriding public interest in free access to harmonised standards, and in response ISO/IEC have asked to annul this decision (Case T-631/24), thereby casting uncertainty regarding the sustainability of the current model for harmonised standards.

    \item\textit{Developing standards with reference to legal requirements and ethical norms:} ISO/IEC and CEN/CENELEC guidelines prohibit addressing legal compliance so as to avoid introducing barriers to trade in these specifications. Harmonised standards address this by including a specific annex to map provisions in the standard to specific legislative requirements. This arms-length treatment and the perceived lack of democratic legitimacy in standards development means that issues of fundamental rights and ethics are limited to non-normative technical reports \cite{noauthor_isoiec_nodate-1}.
  
    \item\textit{Integrating horizontal and vertical standards development:} A key benefit of grounding harmonised standards in international ISO/IEC standards is the support offered for regulatory certification regimes, where conformity assessment is based on the ISO 17000 series of standards. This allows for use of standardised certification assessments from authorised bodies that certify against a given regulation and its harmonised standards through ``notification'' (i.e. competent authorities assess certification providers and ``notify'' them if deemed qualified). Complementing this scheme is the harmonised management system standard (MSS) template defined by ISO/IEC~\cite{noauthor_iso_guide2}. This common template facilitates organisations combining adoption of MSSs from different standards committees into a unified organisational management system, e.g. ISO/IEC 27001 for information security and ISO 42001 which is the MSS developed by SC42 for AI from the harmonised management system. This however offers no guarantee that similar concepts or control measures included from different ISO/IEC standards committees are not in conflict, as they are often developed at different times and as committees addressing horizontal technologies such as AI are typically chartered separately to those addressing sectorial standards, such as ISO/TC 210 on health related products. The ideal solution for a Joint Committee to be formed between separate committees is tricky as there are two sets of consensus forming and voting communities involved which makes it typically slower to establish and complete standards project. To date, the harmonised standards request to JTC21 has only sought horizontal harmonised standards from JTC21, so the development of harmonised standards specific to AI use in a vertical domain such as medical devices may introduce uncertainties in the timeliness and consistency of liaison between these committees on joint specifications.
\end{itemize}

\section{Mapping a Regulatory Learning Space}

\begin{figure*}[ht]
    \centering
   \fbox{\includegraphics[width=\textwidth]{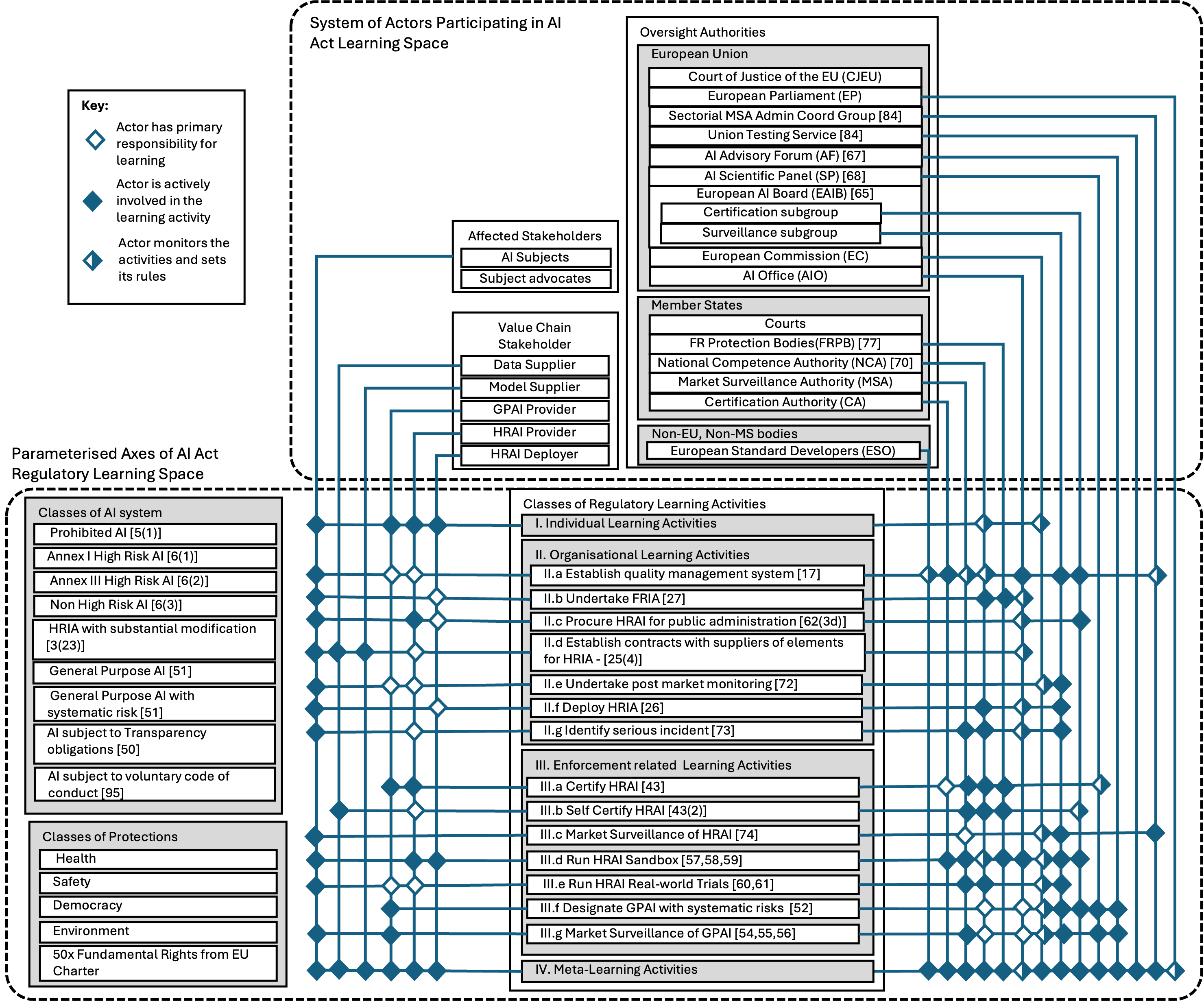}}
    \caption{Parameterised Learning Space for the AI Act}
    \label{fig:learning-space}
\end{figure*}

Regulatory learning is a means by which technology regulation can respond and adapt to rapid advances in the capabilities and use of the technologies subject to regulation. It features in the EU's smart regulation policies \cite{noauthor_communication_2010}, which aim to ensure in general that regulation achieves its policy objectives with the minimum regulatory burden on industry, society and public administration across the legislative lifecycle. The AI Act contains many provisions that invoke a range regulatory learning and revision mechanisms that include sharing and coordination of enforcement actions between the Member State and regulatory bodies (Article 65,66); operating learning environments such a regulatory sandboxes (Article 57, 58, 59) and real-world trials (Article 60); providing and revising supporting technical specification (e.g. harmonised standards, guidelines, codes of practice, codes of conduct or reporting templates); authoritative opinions from expert representative, i.e. advisory forum (Article 67) and scientific panels (Article 68) and modification to the legal code through delegated acts (Article 97), implementing acts (Articles 41, 56, 58, 108, etc.) and periodic review and revision of the legislation (Article 112). 

Regulatory sandboxes have previously found some success with financial regulation \cite{ahern_regulatory_2021} and the AI Act requires at least one in operation in each Member State (Article 57, 58, 59). Sandboxes allow regulators and prospective regulated actors to experience, assess and report on technical measures and their documentation. This is conducted within time and scope-limited constraints set by regulatory bodies to inform the improvement of implementation on both sides. However, as part of a Member State's competence, their scoping and resourcing of sandboxes\footnote{\url{https://artificialintelligenceact.eu/ai-regulatory-sandbox-approaches-eu-member-state-overview/}} may be constrained by national policy priorities, which may result in gaps or duplications across the resulting population of sandboxes. Experiences from finance sector also show the benefits of resolving ambiguities in regulatory requirements and terminology through collaborative learning arenas that do not need to operate under the ambit of regulators \cite{buckley_building_2020}. Mechanisms such as living labs and test-beds can also offer complementary arenas for regulatory learning with different degrees of technology maturity, regulator involvement, resourcing models, access routes and stakeholder engagement~\cite{kert_regulatory_2022} as well as for exploring non-technical governance innovation~\cite{engels_testing_2019}.  Correspondingly, the AI Act includes oversight rules for real-world testing of technical measures using human subjects (Articles 60, 61). However, the degree of autonomy on sandboxes available to AI National Competent Authorities and the potential for valuable regulatory learning from other activities presents a major coordination problem. This falls into the remit of the European AI Board (Article 66) and the EC. Such coordination will need to be handled transparently and systematically to ensure that the many uncertainties raised by the Act across different protections and sectors are addressed in a timely, predictable, efficient and legitimate manner. Further, the risk-based approach adopted from the NLF is at its core an adaptive mechanism for managing uncertainty. Pre-deployment certification aims to ensure that known or reasonably foreseeable risks are sufficiently mitigated within the available state of the art in technical measures. Where new risks materialise or mitigation measures fail in unanticipated ways, the market surveillance system enables such knowledge to be disseminated across the market to rapidly correct or remove affected products and to motivate investment in corresponding mitigation measures. We therefore  view the certification and surveillance activities for implementing the Act (including penalties and appeals to courts) as part of its regulatory learning mechanisms.

These trends in adoption of regulatory learning coupled with the uncertainties in the Act indicate that a broad and structured framework for regulatory learning is required to enable the efficient, timely and stakeholder-inclusive system for resolving uncertainties and adapting to technological advances. This would allow  impacted actors to coordinate to analyse the information sharing requirements between different forms of regulatory learning, to efficiently allocate and schedule private and public resources to learning activities and to minimise wasteful duplication of such activities. To this end, we propose a framework developed from the provisions of the Act and consisting of a \textit{taxonomy of actors} who will participate in learning activities that are organised into a regulatory learning space where the specific learning activities are situated by \textit{classes of protections} being addressed, the \textit{types of AI systems} involved and \textit{classes of learning activities}.  

The classification of regulatory learning actors contains \textit{Value-chain Actors}, \textit{Oversight Authorities} and \textit{Affected Stakeholders}. The class of Value-chain Actors consists of operators of AI systems (defined in Article 3(8) to encompass providers, deployers, product manufacturers and their authorised representatives, distributors and importers) as well as the suppliers of data, AI models and other components with compliance implications (Article 25). The class of Oversight Authorities consists of bodies and agencies that make decisions on the policy lifecycle for the Act. They are classified as EU bodies, Member State bodies and those operating independently of the EU and Member States, such as ESOs. It includes therefore those involved in investigating and making decisions on compliance and in the development of official guidelines, regulatory and legal opinions, delegated/implementing acts, or harmonised standards. 
The class of Affected Stakeholders consists of natural persons who benefit or perceive they may benefit from the Act's protections of their health, safety, fundamental rights including democracy, the rule of law and environmental protection. In this, we also include groups and non-governmental organisations advocating for collective interests on protections. 

The efficiency and legitimacy of the AI Act's implementation, given the complex array of the actors involved, will depend on the networks and patterns of information exchange, normative guidance and accountability mechanisms. We have identified that there are several uncertainties in play already that will affect not only what information needs to be communicated between actors, but also the involvement of actors in patterns and triggers for regulatory communications and the learnings they convey, which represents a large complex network for regulatory interactions. Therefore, as visualised in Figure \ref{fig:learning-space}, we propose a structure of three parametrised axes within which the regulatory learning space can be subdivided, actors' involvement can be planned and resourced, and interplay between different forms of regulatory learning can be managed. 

The proposed axes, derived from the provisions of the Act, are:(1) the types of AI system, (2) the types of protections offered and (3) the types of regulatory learning activities in which actors may participate. The \textit{AI System Types} breaks down as: prohibited; High Risk AI (HRAI) classified under any of the 20 vertical sectorial legislative acts in Annex I; HRAI classified under the 8 areas and 24 sub-areas in Annex III; HRAI treated by MSA as non-high-risk (Article 80) under derogations of Article 6(3); deployed HRAI with substantial modifications; GPAI systems; GPAI systems with systematic risks; AI subject to transparency obligations and those subject to voluntary codes of conduct. \textit{Protections} cover those for health, safety, fundamental rights, democracy and the environment. Finally, \textit{regulatory learning activities} are classified as: those undertaken by individuals, primarily related by AI literacy and regulatory competences; organisational learning required of value-chain actors prior to interactions with oversight authorities; regulatory learning resulting from such enforcement interactions and meta-learning that results from observing the outcomes of other learning activities and applying it to developing or revising rules and recommendations to which those activities are subject \cite{kooiman_meta-governance_2009}.  

In Figure \ref{fig:learning-space}, we then map out the types of possible involvement by the different framework actors with different levels of learning activities, thereby laying out the make-up of the learning arenas where these activities will be undertaken and where different actor classes would be involved in learning activities as implied by the Act. This differentiates between actor classes that are merely participants in the activity and those that are responsible for the conduct of the activity, including the future adoption of its learning outcomes by that particular actor. The mapping also differentiates actors that observe that class of learning activity and use the collective learning outcomes to revise the governance arrangements for that class, i.e. to conduct meta-learning. In this way, the set of actors making up the arena in which specific learning activities are undertaken can be identified in order to plan and efficiently support such activities. The arenas for meta-learning are not directly elaborated in this way, since their make-up is more accurately informed by the organisation and enforcement learning activities from which they learn. This mapping is therefore expanded separately in Table \ref{tab:meta-learning}.
By modelling the interactions between actors and activities involved in implementing the AI Act as classes of learners and learning activities, we illuminate how the Act's extensive system of ``smart regulation'' mechanisms will efficiently and legitimately address known uncertainties. 

However, two important issues arise as a result: first, responsibility for defining learning goals for individuals is not well-defined. Though Article 4 of the Act requires AI literacy in personnel of AI operators, how this should be executed through formal courses, experiential learning \cite{madaio_learning_2024} and competency reviews is undefined. The professional competence of oversight authority personnel is already generally defined under the NLF, e.g. as mentioned in the Act for national competent authorities (Article 70(3)) and notifying bodies (Article 31(11)). However, the responsibility to specify these competences that must draw from vertical legislation of Annex I, horizontal AI provisions and standards and fundamental right concerns resides diffused across the EU and Member States (Recital 20). Further, there is no responsibility to promote AI literacy requirements to upstream suppliers. Independent competency definition activities are emerging, e.g. International Association of Privacy Professions\footnote{\url{https://iapp.org/resources/topics/eu-ai-act/}} and at JTC21 in preliminary investigation into ``competence requirements for AI ethicist professionals'' and ``Guidance for upskilling organisations on AI ethics and social concerns'', but its integration with EU and Member States responsibilities is currently unclear. 

Second, regulatory sandboxes are primarily designed to support the learning by regulated parties in order to help manage the risk to their innovation and investment plans in cooperation with regulators \cite{oecd2023sandbox}, and often also need to support Member States industrial policy to justify investment. Therefore, the systematic gathering of evidence to support consensus building for developing the required implementing and delegated acts \cite{novelli_robust_2024} and assessment and review of the Act (Article 112) in the long term will require coordination and integration outcomes from across the broad range of regulatory meta-learning identified in Table \ref{tab:meta-learning}, based on the analysis in \cite{kaizenner_ai_2024}. The table identifies where a useful information flow is likely between instances of the organisational and enforcement learning activities and the meta-learning activities listed. It also indicates where the outcome of those learning activities will place additional requirement on the other learning activities. It can be seen that there is a high chance of such useful information flow between organisational and enforcement learning activities and meta-learning. In particular, information flows to and from most type of enforcement learning to meta-learning activities are likely as there will be cross over where HRAI enforcement informed GPAI related activities and vice versa based on the increasing prevalence of GPAI use in areas treated as HRAI and HRAI oversight serving as a source of evidence on GPAI risks and thereby feedback on their classification. This dense pattern of bidirectional information exchange to and from meta-learning reinforces the need for systematic support for finding, exchanging and meaningfully consuming the outcome of different learning activities in an efficient manner.

\begin{table*}[t!]
\caption{Information flow between organisational and enforcement learning activities and meta learning activities with signs indicating: $\leftarrow$ as instance of learning activity informs meta-learning activity, $\rightarrow$ as outcome of meta-learning activity places requirements on instances of learning activities; $\leftrightarrow$ as both these apply; . (dot) as no interactions; and acronyms as DA for Delegated Act; IA for Implementing Act; and [] (square brackets) indicating AI Act article\label{tab:meta-learning}}

\begin{center}
\begin{tabular}{|m{6cm}||c|c|c|c|c|c|c||c|c|c|c|c|c|c|} 
 \hline

 \multicolumn{1}{|c||}{I.  Meta-learning activities}& 

 \multicolumn{7}{|c||}{II. Organisational learning activities} & \multicolumn{7}{|c|}{III. Enforcement learning activities}\\
 \hline
& a & b & c & d & e & f & g & a & b & c & d & e & f & g\\ 
 \hline
 \hline
 a: Develop harmonised standards [40]& $\leftrightarrow$ & $\leftrightarrow$ & $\leftrightarrow$& $\leftrightarrow$ & $\leftrightarrow$ & $\leftrightarrow$& $\leftrightarrow$ & $\leftrightarrow$ & $\leftrightarrow$ & $\leftrightarrow$ & $\leftrightarrow$ & $\leftrightarrow$ & $\leftarrow$ & $\leftarrow$ \\
 \hline
 b: Revise Anx III derogations [6,97] DA& $\leftrightarrow$ & $\leftrightarrow$ & $\leftrightarrow$ & . & $\leftarrow$ & . & $\leftrightarrow$ & $\leftrightarrow$ & $\leftrightarrow$ & $\leftrightarrow$ & $\leftrightarrow$ & $\leftrightarrow$ & $\leftrightarrow$ & $\leftrightarrow$ \\
 \hline
 c: Revise Anx I and III [7,97] DA& $\rightarrow$ & $\leftarrow$ & $\leftrightarrow$ & . & $\leftarrow$ & . & $\leftarrow$ & $\leftarrow$ & $\leftarrow$ & $\leftarrow$ & $\leftrightarrow$ & $\leftrightarrow$ & $\leftarrow$ & $\leftarrow$ \\
 \hline
 d: Revise technical documentation spec Anx IV [6,97] DA& $\leftrightarrow$ & $\leftarrow$ & $\leftrightarrow$ & $\leftrightarrow$ & $\leftrightarrow$ & $\leftrightarrow$ & $\leftrightarrow$ & $\leftrightarrow$ & $\leftrightarrow$ & $\leftrightarrow$ & $\leftrightarrow$ & $\leftrightarrow$ & $\leftarrow$ & $\leftarrow$ \\
 \hline
 e: Revise conformity assessment Anx VI and VII [43,97] DA& $\leftrightarrow$ & $\leftarrow$ & $\leftrightarrow$ & $\leftrightarrow$ & $\leftrightarrow$ & . & $\leftarrow$ & $\leftrightarrow$ & $\leftrightarrow$ & $\leftrightarrow$ & $\leftrightarrow$ & $\leftrightarrow$ & $\rightarrow$ & $\leftrightarrow$ \\
 \hline
 f: Revise declaration of conformity spec Anx V [47,97] DA& $\leftrightarrow$ & $\leftarrow$ & $\leftrightarrow$ & $\leftrightarrow$ & $\leftrightarrow$ & $\leftarrow$ & . & $\leftrightarrow$ & $\leftrightarrow$ & $\leftrightarrow$ & $\leftrightarrow$ & $\leftrightarrow$ & $\rightarrow$ & $\rightarrow$ \\
 \hline
 g: Revise GPAI threshold and criteria in Anx XIII [52,97] DA& $\leftrightarrow$ & $\leftrightarrow$ & $\leftarrow$ & $\rightarrow$ & . & . & $\leftarrow$ & $\leftarrow$ & $\leftarrow$ & $\leftarrow$ & $\leftarrow$ & $\leftarrow$ & $\leftrightarrow$ & $\leftrightarrow$   \\
 \hline
 h: Revise GPAI tech document Anx XI and downstream in Anx XII [53,97] DA& $\leftrightarrow$ & $\leftarrow$ & $\leftrightarrow$ & $\leftrightarrow$ & $\leftrightarrow$ & $\leftrightarrow$ & $\leftrightarrow$ & $\leftrightarrow$ & $\leftrightarrow$ & $\leftrightarrow$ & $\leftrightarrow$ & $\leftrightarrow$ & $\leftrightarrow$ & $\leftrightarrow$  \\
 \hline
 i: Develop common specification [41,98] IA& $\leftrightarrow$ & $\leftrightarrow$ & $\leftrightarrow$ & $\leftrightarrow$ & $\leftrightarrow$& $\leftrightarrow$ & $\leftrightarrow$ & $\leftrightarrow$ & $\leftrightarrow$ & $\leftrightarrow$ & $\leftrightarrow$ & $\leftrightarrow$ & $\leftarrow$ & $\leftarrow$ \\
 \hline
 j: Develop spec for reg sandboxes  [58,98] IA& $\leftarrow$ & $\leftarrow$ & $\leftarrow$ & $\leftarrow$ & $\leftarrow$ & $\leftarrow$ & $\leftarrow$ & $\leftarrow$ & $\leftarrow$ & $\leftarrow$ & $\leftrightarrow$ & $\leftarrow$ & $\leftarrow$ & $\leftarrow$  \\
 \hline
 k: Develop spec for real-world testing [60,98] IA& $\leftarrow$ & $\leftarrow$ & $\leftarrow$ & $\leftarrow$ & $\leftarrow$ & $\leftrightarrow$ & $\leftrightarrow$ & $\leftarrow$ & $\leftarrow$ & $\leftarrow$ & $\leftarrow$ & $\leftrightarrow$ & $\leftarrow$ & $\leftarrow$ \\
 \hline
 l: Develop template for post market monitoring plans [72,98] IA& $\leftarrow$ & $\leftarrow$ & $\leftarrow$ & $\leftarrow$ & $\leftrightarrow$ & $\leftrightarrow$ & $\leftrightarrow$ & $\leftrightarrow$ & $\leftrightarrow$ & $\leftrightarrow$ & $\leftrightarrow$ & $\leftrightarrow$ & $\leftarrow$ & $\leftrightarrow$ \\
 \hline
 m: Develop rule for GPAI evaluation [92,98] IA& $\leftarrow$ & $\leftarrow$ & $\leftarrow$ & $\leftarrow$ & $\leftarrow$ & $\leftarrow$ & $\leftarrow$ & $\leftarrow$ & $\leftarrow$ & $\leftarrow$ & $\leftarrow$ & $\leftarrow$ & $\leftrightarrow$ & $\leftrightarrow$  \\
 \hline
 n: Develop criteria non-high-risk Anx III systems [6]& $\leftrightarrow$ & $\leftrightarrow$ & $\leftarrow$ & . & . & . & $\leftarrow$ & $\leftrightarrow$ & $\leftrightarrow$ & $\leftrightarrow$ & $\leftrightarrow$ & $\leftrightarrow$ & $\leftrightarrow$ & $\leftrightarrow$ \\
 \hline
 o: Develop QMS guidelines for microenterprises [63] & $\leftrightarrow$ & $\leftrightarrow$ & $\leftarrow$ & $\leftrightarrow$ & $\leftrightarrow$ & $\rightarrow$ & $\leftrightarrow$ & $\leftrightarrow$ & $\leftrightarrow$ & $\leftrightarrow$ & $\leftrightarrow$ & $\leftrightarrow$ & . & .  \\
 \hline
 p: Develop guidelines HRAI compliance implementation [96]  & $\leftrightarrow$ & $\leftrightarrow$ & $\leftrightarrow$ & $\leftrightarrow$ & $\leftrightarrow$ & $\leftrightarrow$ & $\leftrightarrow$ & $\leftrightarrow$ & $\leftrightarrow$ & $\leftrightarrow$ & $\leftrightarrow$ & $\leftrightarrow$ & $\leftarrow$ & $\leftarrow$  \\
 \hline
 q: Develop template for tech document by small and micro enterprises [11]& $\leftrightarrow$ & $\leftarrow$ & $\leftrightarrow$ & $\leftrightarrow$ & $\leftrightarrow$ & $\rightarrow$ & $\leftrightarrow$ & $\leftrightarrow$ & $\leftrightarrow$ & $\leftrightarrow$ & $\leftrightarrow$ & $\leftrightarrow$ & $\leftarrow$ & $\leftrightarrow$ \\
 \hline
 r: Develop model contract terms for AIP and suppliers [25] & $\leftrightarrow$ & $\leftarrow$ & $\leftarrow$ & $\leftrightarrow$ & $\leftrightarrow$ & . & $\leftarrow$ & $\leftrightarrow$ & $\leftrightarrow$ & $\leftrightarrow$ & $\leftrightarrow$ & $\leftrightarrow$ & $\leftrightarrow$ & $\leftrightarrow$  \\
 \hline
 s: Encourage development of benchmarks and development methodologies [15(2)] & $\leftrightarrow$ & $\leftrightarrow$& $\leftrightarrow$ & $\leftarrow$ & $\leftrightarrow$ & $\rightarrow$ & $\leftrightarrow$ & $\leftrightarrow$ & $\leftrightarrow$ & $\leftrightarrow$ & $\leftrightarrow$ & $\leftrightarrow$ & $\leftrightarrow$ & $\leftrightarrow$ \\
 \hline
 t: Develop template for GPAI training copyright report [53(1d)] & $\leftrightarrow$ & $\leftrightarrow$ & $\leftarrow$ & $\leftrightarrow$ & . & $\rightarrow$ & $\leftarrow$ & . & $\leftrightarrow$ & $\leftrightarrow$ & $\leftrightarrow$ & $\leftarrow$ & $\leftrightarrow$ & $\leftrightarrow$  \\
 \hline
 u: Develop objective and participative methodology for the evaluation of risk levels [112(11)]  & $\leftrightarrow$ & $\leftrightarrow$ & $\leftrightarrow$ & $\leftrightarrow$ & $\leftrightarrow$& $\leftrightarrow$ & $\leftrightarrow$ &$\leftrightarrow$ & $\leftrightarrow$ & $\leftrightarrow$ & $\leftrightarrow$ & $\leftrightarrow$ & $\leftrightarrow$ & $\leftrightarrow$  \\
 \hline
 v: Develop GPAI code of practice [56]  & $\leftrightarrow$ & $\leftrightarrow$ & $\leftrightarrow$ & $\leftrightarrow$ & $\leftrightarrow$ & $\leftrightarrow$ & $\leftrightarrow$ & $\leftrightarrow$ & $\leftrightarrow$ & $\leftrightarrow$ & $\leftrightarrow$ & $\leftrightarrow$ & $\leftrightarrow$ & $\leftrightarrow$ \\
 \hline
 w: Revise AI Act [112]  & $\leftrightarrow$ & $\leftrightarrow$ & $\leftrightarrow$ & $\leftrightarrow$ & $\leftrightarrow$ & $\leftrightarrow$ & $\leftrightarrow$ & $\leftrightarrow$ & $\leftrightarrow$ & $\leftrightarrow$ & $\leftrightarrow$ & $\leftrightarrow$ & $\leftrightarrow$ & $\leftrightarrow$  \\
 \hline 
\end{tabular}
\end{center}
\end{table*}

It can also be seen that involvement of affected stakeholders is a widespread expectation across activities, prompted by their importance in understanding rights protection and the legitimacy their involvement may offer. However, they are offered relatively little formal role in the Act beyond the right to lodge a complaint with an MSA on rights protection breach (Article 85), to request an explanation of an AI decision (Article 86) and to benefit from whistle-blower protections (Article 87). However, structuring their participation in the design of technical and governance measures is challenging to implement \cite{groves_going_2023}.

\section{Learning via Information Exchange}



Any attempt to ground technology ethics in co-regulation with democratic legitimacy requires compliance to be achieved in a stable manner and implemented efficiently \cite{ferretti_institutionalist_2022}. As we discussed earlier, the AI Act currently presents major legal uncertainties, especially in the effectiveness of protections across the broad range of fundamental rights and in how its horizontal provisions will integrate with vertical enforcement practices of the NLF. While not addressing these uncertainties directly, the Act does already contain a wide range of measures to update its provision in response to technological advances and improved evidence on risks. We do not aim to predict changes arising from regulatory learning, but instead offer a structured framework for considering the Act's provisions as a system of interacting regulatory learning activities. In this way, we can consider how such learnings can be supported in a predictable and efficient manner. In particular, we now consider how information exchange between different learning activities can support efficiency and legitimacy. Efficiency may be derived by improving the semantic and syntactic interoperability of information representing the context and outcomes of learning activities. This not only supports exchange of such information between AI operators and oversight authorities, but with suppliers upstream in the value chain. Legitimacy can be derived from adopting open standards for such information exchange and doing so via platforms that maximise transparency for affected stakeholders and other operators, with the necessary restrictions of commercial confidentiality. Standardised information and structured information-sharing practices may enable affected stakeholders, their advocates, and independent researchers to make public-interest comparisons—such as evaluating the outcomes of individual AI training processes or assessing the appropriateness of thresholds used to measure specific types of harm during system testing.

With this view, we consider the AI Act as also advocating the collection and use of specific knowledge to both determine and ensure compliance, as well as to promote a culture of organisational oversight through the use of this information. An example of this is Articles 9 to 15 where specific categories of information are required to be maintained and used towards indicated processes. However, given that the AI Act ecosystem (or the AI value chain) will mostly consist of multiple entities who develop, provide, and use AI systems, and based on the application of the learning loops from before, we can see how the information required to be maintained by the AI Act will also be required to be shared with other stakeholders outside of the immediate organisational context. At the same time, the specific nature of this information (e.g. technical or legal) also shows challenges in effectively utilising it within the organisation where there will be different levels of expertise and the decision-making person may not always be capable of understanding it.


Thus, we point out that effective enforcement of the AI Act depends on the information sharing of relevant stakeholders, and that doing so efficiently is a necessary regulatory learning in this space. However, the primary challenge in developed shared information modelling approaches is the divergence in use-cases and requirements, particularly across sectors. For addressing this, we require a \textit{common foundational ontology} aligned with the AI Act itself that provides common terminology and basis for developing further approaches that can be provisional while the AI Act is being initially interpreted, and then progressively be extended as certainty of regulatory interpretation is achieved. For such approaches developed in parallel across use-cases, it would be desirable to share information by `aligning' or `mapping' the information so that it can be `translated' and utilised conveniently.

To achieve all of this, we believe the best solution is the use of standards-based information models and processes developed to share them -- specifically \textit{semantic web standards} \cite{DCAT} and their use in data portals that enable cataloguing information and deploying a shared infrastructure that promotes data availability and leads to innovative solutions based on its reuse. The EU has successfully utilised this approach in its Open Data Portal which federates datasets from national and regional portals in Member States. The EU's vision of data spaces is grounded in a similar argument which entirely relies on shared information and processes to achieve greater efficiency through consistency in shared contexts.
The EU Commission's European Interoperability Framework (EIF) 
was developed with a similar goal for creating interoperable digital public services. We posit that a similar approach for the AI value chain, based on a focus on achieving the necessary knowledge exchange for AI Act compliance, should be considered an essential part of regulatory learning in this space.
Another example is the EU's eProcurement Ontology \cite{eProcurementOntology} 
which provides a standardised ontology of concepts for interoperable modelling of procurement data and processes based on EU regulations and directives. The ontology caters to common terminology between stakeholders, and enables the development of common tooling, platforms, and services that function both before and after the award. A similar effort for the implementation of the AI Act can be envisioned that functions `end-to-end' across the value chain and assists in the definition of interoperable information for facilitating sharing between stakeholders, and to formalise the exchange of knowledge necessary for regulatory learning within and across the defined learning loops. More importantly, it can function as the common technical foundation for the implementation of future regulations and directives which will build on top of the AI Act in both horizontal (general) and vertical (sectorial) directions. We point to examples of feasible and sustainable community efforts where such use of semantic web standards towards legal knowledge modelling has been successful in adapting to evolving regulations and standards \cite{pandit_implementing_2024,j_pandit_data_2025}.

Thus, we argue that the development of an interoperable framework is an investment in the regulation and will aid regulatory learning processes for both organisations and policy makers by giving them a common point of reference for developing further obligations and smoothing over reporting obligations. This type of `harmonisation' is critically required given the recent Draghi report 
outlined reduction in regulatory reporting as one of the key factors essential to ensuring EU competitiveness, and where the Commission's response agrees with this view as it has tabled an `omnibus regulation' in its agenda for the 2025 work program that reduces reporting and red-tape by 25\%. 
We interpret this as the Commission aiming to make regulatory enforcement more efficient rather than diluting the regulatory obligations and protection of fundamental rights and freedoms. Thus, our argument for using regulatory learning as a tool to systematise compliance processes across the ecosystem and across regulations should be considered an important factor for such future policy-making decisions. 

To achieve this, we envision the reuse of existing standardisation processes, such as in ISO/IEC, CEN/CENELEC, and W3C to develop appropriate agreements on information exchange formats and processes based on using them. This also provides a way for the integration of information required or involved in the use of harmonised standards for the AI Act within the proposed interoperability framework.



\section{Discussion and Further Work}

We argue that the technology pacing problem coupled with uncertainties related to fundamental rights protections and integration of horizontal provisions and vertical enforcement requires a systematic approach to regulatory learning for the AI Act. We therefore proposed a conceptual, parametrised space for locating instances of learning activities to aim in their planning, resourcing, coordination and information exchange. To ensure regulatory learning is timely, efficient and legitimate we recommend: 
\begin{enumerate}
    \item The development of open, standards-based and machine readable information models for the exchange of the context and learning outcomes of learning activities;
    \item Maximising opportunities for sharing instance of such learning outcomes according to FAIR principles, e.g. accompanying declaration of conformity used in public administration in an EU database;
    \item Where confidentiality restricts open sharing of learning outcomes, encourage the use of data spaces for controlled access to information sharing. 
\end{enumerate}

Considering the AI Act as a learning framework in this way may portend a ``regulatory turn'' in the field of AI ethics, where questions focus on the efficient operation of such horizontal regulation and its effectiveness in protecting fundamental rights. In the context of the Act, such a regulatory turn in AI ethics may refocus voluntary codes more specifically on fundamental rights protections in order to leverage the learning and evidence that will arise from the compliance actions for high-risk and GPAI systems. Such a regulatory turn may have import beyond the EU both due to the extraterritorial impact on international AI value chains known as the ``Brussels effect'', and the widespread comparable legal protections in place across different jurisdictions for fundamental, or human, rights and corresponding legitimacy in comparison to proposed ethical frameworks for AI \cite{lewis_global_2020}.

While our analysis is limited to the scope of the AI Act, given the breadth of recent EU digital regulations, the next step is to explore regulatory learning systems that can address the interplay of the Act and related digital and product legislations. As discussed above, there is coordination of the interplay between the AI Act and GDPR and also some options for coordination with Annex I product legislation under NLF. However, there is potential interplay with other EU law where their interactions with the Act have not yet received close attention - such as on the cybersecurity of AI (Article 15) with the Cyber Resilience Act 2024/2847 and the Network and Information Systems (NIS2) Directive 2022/2555. Another avenue for exploring coordination involves the system for accrediting trusted research used in the Digital Services Act 2022/2065 to be mobilised for researchers to advance AI and data auditing techniques with HRAI and GPAI providers. More broadly, effective implementation of AI quality management systems and risk assessments, especially for emerging impact on fundamental rights can also feed in fruitfully for provider obligations under liability legislation including the General Product Safety Regulation 2023/988, the Product Liability Directive 2024/2853, class actions via the Representative Action Directive 2020/1828, and Corporate Sustainability Reporting Directive 2022/2464, as well as the Corporate Sustainability Due Diligence Directive 2024/1760  on environmental and social reporting. The European Data Protection Supervisor has recently proposed a digital clearing house to facilitate multilateral cooperation between authorities involved in enforcing the EU digital aquis \cite{epds_towards_2025}. Expanding our learning space model as an open data resource may offer a useful resource for such coordination but also a mean for engaging  affected stakeholder and civil society on these complex issues.

The scope of the analysis of information exchange could be expanded further to include AI risk and compliance frameworks as they emerge outside of the EU jurisdiction such as the US, China, Canada, UK and Singapore. Sandboxes are already being adopted internationally as these other jurisdictions grapple with similar AI risk uncertainties. 
Analysing the potential for learning interchange between sandboxes across jurisdictions may help assess where enforcement of one measure already adequately offers outcomes intended by the other or where minor incompatibilities could be readily addressed by bodies tasked with addressing issues raised as part of international trade agreements.

Finally, the profound impacts AI seems likely to bring to many aspects of our lives may result in the enforcement of the AI Act exposing new vectors where different fundamental rights come into conflict in unanticipated ways. Insights may arise from cases brought before national courts or the CJEU, or where a MSA or the AI Office makes a ruling based on an identified fundamental rights breach. Here, learning could be facilitated by cataloguing AI fundamental rights opinions from these different types of learning together, contrasting with the more siloed public catalogues related to existing rights protecting frameworks such as GDPR Supervisory Authorities and the ECtHR. This may support accelerated legal scholarship on topics such as clarification of circumstances under which the AI Act does or does not support fundamental rights protections and understanding where different protections may come into conflict (including the freedom to conduct a business). 

\section{Acknowledgments}

This project has received funding from the European Commission's Horizon Europe Research and Innovation Programme under grant agreement No. 101177579 (FORSEE) and from the ADAPT Centre for Digital Media Technology, which is funded by Research Ireland and is co-funded under the European Regional Development Fund (ERDF) through Grant\#13/RC/2106\_P2.

\bibliography{aaai25}

\end{document}